\documentclass[prl,amsmath,amssymb,twocolumn]{revtex4-1}

\usepackage[english]{babel}
\usepackage{amsmath,amssymb,hyperref,bbm,graphicx,color,txfonts}

\renewcommand{\Re}{\mathop{\mathrm{Re}}}
\renewcommand{\Im}{\mathop{\mathrm{Im}}}
\newcommand{\abs}[1]{\left\lvert #1 \right\rvert}

\newcommand{\be}{\begin{equation}}
\newcommand{\beq}{\begin{eqnarray}}
\newcommand{\eeq}{\end{eqnarray}}

\def \be{\begin{equation}}
\def \ee{\end{equation}}
\def \ba{\begin{array}}
\def \ea{\end{array}}
\def \bea{\begin{eqnarray}}
\def \eea{\end{eqnarray}}

\def \br{{\bf r}}
\def \bs{{\bf s}}

\def \L{{\Lambda}}

\def \b{{\beta}}
\def \g{{\gamma}}

\def \w{{\omega}}

\def \G{{\Gamma}}

\def \av#1{{\langle#1\rangle}}

\begin{document}

\title{Dynamical Critical Phenomena in Driven-Dissipative Systems}
\author{L. M. Sieberer$^{1,2}$, S. D. Huber$^{3,4}$, E. Altman$^{4,5}$, and S. Diehl$^{1,2}$
\\{$^1$\small \em  Institute for Theoretical Physics, University of Innsbruck, A-6020 Innsbruck, Austria}\\
{$^2$\small \em Institute for Quantum Optics and Quantum Information of the Austrian Academy of Sciences, A-6020 Innsbruck, Austria}\\
{$^3$\small \em Theoretische Physik, Wolfgang-Pauli-Strasse 27, ETH Zurich, CH-8093 Zurich, Switzerland}\\
{$^4$\small \em Department of Condensed Matter Physics, Weizmann Institute of Science, Rehovot 76100, Israel}\\
{$^5$\small \em Department of Physics, University of California, Berkeley, CA 94720, USA}\\
}
\begin{abstract}
  We explore the nature of the Bose condensation transition in driven open
  quantum systems, such as exciton-polariton condensates. Using a functional
  renormalization group approach formulated in the Keldysh framework, we
  characterize the dynamical critical behavior that governs decoherence and an
  effective thermalization of the low frequency dynamics. We identify a critical
  exponent special to the driven system, showing that it defines a new dynamical
  universality class. Hence critical points in driven systems lie beyond the
  standard classification of equilibrium dynamical phase transitions. We show
  how the new critical exponent can be probed in experiments with driven cold
  atomic systems and exciton-polariton condensates.
\end{abstract}
\maketitle

Recent years have seen major advances in the exploration of many-body systems in
which matter is strongly coupled to light~\cite{carusotto12}. Such systems
include for example polariton condensates~\cite{kasprzak06}, superconducting
circuits coupled to microwave resonators~\cite{schoelkopf08,clarke08}, cavity
quantum electrodynamics~\cite{hartmann08} as well as ultracold atoms coupled to
high finesse optical cavities~\cite{ritsch12}.  As in traditional quantum optics
settings, these experiments are subject to losses, which may be compensated by
continuous drive, yet they retain the many-body character of condensed
matter. This combination of ingredients from atomic physics and quantum optics
in a many-body context defines a qualitatively new class of quantum matter far
from thermal equilibrium.  An intriguing question from the theoretical
perspective is what new universal behavior can emerge under such conditions.

A case in point are exciton-polariton condensates. Polaritons are short lived
optical excitations in semiconductor quantum wells. Continuous
pumping is required to maintain their population in steady state. But in spite
of the non-equilibrium conditions, experiments have
demonstrated Bose condensation~\cite{kasprzak06} and, more recently,
 have even observed the establishment of a critical phase with
power-law correlations in a two dimensional system below a presumed
Kosterlitz-Thouless phase transition~\cite{roumpos12}. At a
fundamental level however there is no understanding of the condensation
transition in the presence of loss and external drive, and more generally of
continuous phase transitions under such conditions.

In this letter we develop a theory of dynamical critical phenomena in
driven-dissipative systems in three dimensions.  Motivated by the experiments
described above we focus on the case of Bose condensation with the following key
results. \emph{(i) Low-frequency thermalization} -- The microscopic dynamics of
a driven system is incompatible with an equilibrium-like Gibbs distribution at
steady state. Nevertheless a scale independent effective temperature emerges at
low frequencies in the universal regime near the critical point, and all
correlations in this regime obey a classical fluctuation-dissipation relation
(FDR). Such a phenomenon of low frequency effective equilibrium has been
identified previously in different
contexts~\cite{mitra06,diehl08,dallatorre10,dallatorre12,oztop12,wouters2006}. \emph{(ii)
  Universal low-frequency decoherence} -- In spite of the effective
thermalization, the critical dynamics is significantly affected by the
non-equilibrium conditions set by the microscopic theory. Specifically we show
that all coherent dynamics, as measured by standard response functions, fades
out at long wavelengths as a power-law with a new universal critical
exponent. The decoherence exponent cannot be mimicked by any equilibrium model
and places the critical dynamics of a driven system in a new dynamical
universality class beyond the Halperin-Hohenberg classification of equilibrium
dynamical critical behavior~\cite{hohenberg77}.

\emph{Open system dynamics--} A microscopic description of driven open systems
typically starts from a Markovian quantum master equation or an equivalent
Keldysh action (see Supplementary Information (SI)). However, the novel aspects
in the critical dynamics of driven dissipative systems discussed below can be
most simply illustrated by considering an effective mesoscopic description of
the order parameter dynamics using a stochastic Gross-Pitaevskii equation
\cite{carusotto2005}
\begin{equation}
  \label{SGP}
  i \partial_t \psi = \left[ - \left( A - i D \right) \nabla^2 - \mu + i \chi +
    \left( \lambda - i \kappa \right) \abs{\psi}^2 \right] \psi
  + \zeta.
\end{equation}
As we show below, this equation can be rigorously derived from a fully quantum
microscopic description of the condensate when including only the relevant terms
near the critical point. The different terms in~\eqref{SGP} have a clear
physical origin. $ \chi= \left( \g_p - \g_l \right)/2$ is the effective gain,
which combines the incoherent pump field minus the local single-particle loss
terms. $\kappa,\lambda>0$ are respectively two-body loss and interaction
parameters. The diffusion term $D$ is not contained in the original microscopic
model, and is not crucial to describe most non-universal aspects of, e.g.,
exciton-polariton condensates \cite{wouters07} (but see~\cite{wouters10}). In a
systematic treatment of long-wavelength universal critical behavior, however,
such term is generated upon integrating out high frequency modes during the
renormalization group (RG) flow, irrespective of its microscopic value. We
therefore include it at the mesoscopic level with a phenomenological
coefficient. Finally $\zeta$ is a Gaussian white noise with correlations
$\av{\zeta^*(t,\mathbf{x}) \zeta(t',\mathbf{x}')} = \g\delta(t-t')
\delta(\mathbf{x} - \mathbf{x}') $ where $\gamma = \gamma_p + \gamma_l$. Such
noise is necessarily induced by the losses and sudden appearances of particles
due to pumping.

The dGP describes a mean field transition from a stationary
condensate solution with density $|\psi|^2=\chi/\kappa$ for $\chi
> 0$ to the vacuum state when $\chi$ crosses zero. Dynamical
stability~\cite{keeling10} determines the chemical potential as
$\mu=\lambda|\psi|^2$. Similar to a temperature, the noise term
in Eq.~\eqref{SGP} can drive a transition at finite particle
density, thereby inducing critical fluctuations.

As the equation of motion is cast in Langevin form, one might suspect that it
can be categorized into one of the well-known models of dynamical critical
phenomena classified by Hohenberg and Halperin~\cite{hohenberg77}. However, this
is not true in general. Crucially coherent (real parts of the couplings in
Eq.~\eqref{SGP}) and dissipative (imaginary parts) dynamics have different
physical origins in driven-dissipative systems. In particular, the dissipative
dynamics is determined by the intensity of the pump and loss terms,
independently of the intrinsic Hamiltonian dynamics of the system. Equilibrium
models~\cite{hohenberg77}, on the other hand, are constrained to have a specific
relation between the reversible and dissipative terms to ensure a thermal Gibbs
ensemble in steady state~\cite{chaikin95:_princ,tauber07} (see below). The
unconstrained dynamics in driven systems is the key feature that can lead to
novel dynamic critical behavior.

\emph{Microscopic Model} -- Having illustrated the nature of the problem with
the effective classical equation~\eqref{SGP} we turn to a fully quantum
description within the Keldysh framework. Our starting point is a non-unitary
quantum evolution described by a many-body master equation in Lindblad form, or
equivalently by the following dissipative Keldysh action (see SI for details of
the correspondence)
\begin{multline}
  \mathcal{S} = \int_{t,\mathbf{x}} \biggl\{ \left( \phi_c^{*},\phi_q^{*}
\right)
    \begin{pmatrix}
      0 & P^A\\
      P^R & P^K
    \end{pmatrix}
    \begin{pmatrix}
      \phi_c \\ \phi_q
    \end{pmatrix}  + i 4 \kappa \phi_c^{*} \phi_c \phi_q^{*} \phi_q \\
    - \left[ \left( \lambda + i \kappa \right) \left( \phi_c^{*2} \phi_c \phi_q
        + \phi_q^{*2} \phi_c \phi_q \right) + c.c.  \right] \biggr\}.
    \label{eq:micro}
\end{multline}
Here $\phi_c$, $\phi_q$ are the "classical" and "quantum" fields, defined by the
symmetric and anti-symmetric combinations of the fields on the forward and
backward parts of the Keldysh contour~\cite{kamenev09:_keldy,altlandsimons}. The
microscopic inverse Green's functions are given by $P^R = i \partial_t + A
\nabla^2 + \mu - i \chi$, $P^A = P^{R \dag}$, $P^K = i \gamma$.

The importance of the various terms in the microscopic action~\eqref{eq:micro}
in the vicinity of the critical point can be inferred from canonical power
counting, which serves as a valuable guideline for the explicit evaluation of
the problem. Vanishing of the mass scale $\chi$ defines a Gaussian fixed point
with dynamical critical exponent $z=2$ ($ \w\sim k^z$, $k$ a momentum
scale). Canonical power counting determines the scaling dimensions of the fields
and interaction constants with respect to this fixed point: At criticality, the
spectral components of the Gaussian action scale as $P^{R/A} \sim k^2$, while
the Keldysh component generically takes a constant value, i.e., $P^K \sim
k^0$. Hence, to maintain scale invariance of the quadratic action, the scaling
dimensions of the fields must be $[\phi_c] = \frac{d - 2}{2}$ and $[\phi_q] =
\frac{d + 2}{2}$. From this result we read off the canonical scaling dimensions
of the interaction constants. This analysis shows that in the case of interest
$d=3$, local vertices containing more than two quantum fields or more than five
classical fields are irrelevant. For the critical problem, the last terms in
both lines of Eq.~\eqref{eq:micro} can thus be skipped, massively simplifying
the complexity of the problem. The only marginal term with two quantum fields is
the Keldysh component of the single-particle inverse Green's function, i.e., the
noise vertex. In this sense, the critical theory is equivalent to a stochastic
\emph{classical} problem~\cite{msr73,dedominics76}, as previously observed
in~\cite{mitra06,mitra11_1}. But as noted above it cannot be \emph{a priori}
categorized in one of the dynamical universality classes~\cite{hohenberg77}
subject to an intrinsic equilibrium constraint.

\emph{Functional RG} -- In order to focus quantitatively on the critical
behavior we use a functional RG approach formulated originally by
Wetterich~\cite{wetterich93} and adapted to the Keldysh real time framework in
Refs.~\cite{gasenzer08,berges09} (see SI for details). At the formal level this
technique provides an exact functional flow equation for an effective action
functional $\G_\L[\phi_c,\phi_q]$, which includes information on increasingly
long wavelength fluctuations (at the microscopic cutoff scale $\G_{\L_0}\approx
\mathcal S$). In practice one works with an ansatz for the effective action and
thereby projects the functional flow onto scaling equations for a finite set of
coupling constants. For the description of general
equilibrium~\cite{berges02,salmhofer01,pawlowski07,delamotte07,rosten12,boettcher12}
and Ising dynamical~\cite{canet07} critical behavior the functional RG gave
results that are competitive with high-order epsilon expansion and with Monte
Carlo simulations already in rather simple approximation schemes.

Our ansatz for the effective action is motivated by the power counting arguments
introduced above. We include in $\G_\L$ all couplings that are relevant or
marginal in this scheme:
\begin{widetext}
  \begin{equation}
  \label{eq:ansatz}
  \Gamma_{\Lambda} = \int_{t,\mathbf{x}} \left\{ \left( \phi_c^{*},\phi_q^{*} \right)
      \begin{pmatrix}
        0 & i Z \partial_t + \bar{K} \nabla^2 \\
        i Z^{*} \partial_t + \bar{K}^{*} \nabla^2 & i \bar{\gamma}
      \end{pmatrix}
      \begin{pmatrix}
        \phi_c \\ \phi_q
      \end{pmatrix} - \left( \frac{\partial \bar{U}}{\partial \phi_c} \phi_q +
        \frac{\partial \bar{U}^{*}}{\partial \phi_c^{*}} \phi_q^{*} \right)
    \right\}.
  \end{equation}
\end{widetext}  
The dynamical couplings $Z$ and $\bar K$ have to be taken complex valued in
order to be consistent with power counting, even if the respective imaginary
parts vanish (or are very small) at the microscopic scale: Successive momentum
mode elimination implemented by the RG flow generates these terms due to the
simultaneous presence of local coherent and dissipative couplings in the
microscopic model. The fact that the spectral components of the effective action
depend only linearly on $\phi_q$ allowed us to introduce an effective potential
$\bar U$ determined by the complex static couplings. $\bar{U}(\rho_c)=
\frac{1}{2} \bar{u} \left( \rho_c - \rho_{0} \right)^2 + \frac{1}{6} \bar{u}'
\left( \rho_c - \rho_{0} \right)^3$ is a function of the $U(1)$ invariant
combination of classical fields $\rho_c = \phi^*_c \phi_c$ alone. It has a
mexican hat structure ensuring dynamical stability. With this choice we approach
the transition from the ordered side, taking the limit of the stationary state
condensate $\rho_0 = \phi_c^* \phi_c^{} |_{\rm ss} = \phi_0^* \phi_0^{} \to 0$.

All the parameters appearing in~\eqref{eq:ansatz} including the stationary
condensate density $\rho_0$ are functions of the running cutoff
$\Lambda$. Hence, the functional flow of $\Gamma_{\Lambda}$ is reduced by means
of the approximate ansatz to the flow of a finite number of couplings
$\mathbf{g} = \left( Z, \bar{K}, \rho_0, \bar{u}, \bar{u}', \bar{\gamma}
\right)^T$ determined by the $\beta$-functions $\Lambda \partial_{\Lambda}
\mathbf{g} = \beta_{\mathbf{g}}(\mathbf{g})$ (see SI). The critical system is
described by a scaling solution to these flow equations. It is obtained as a
fixed point of the flow of dimensionless renormalized couplings, which we derive
in the following. First we rescale couplings with $Z$,
\begin{equation}
  \label{eq:1}
  K = \bar{K}/Z, \quad u = \bar{u}/Z, \quad u' = \bar{u}'/Z, \quad \gamma =
  \bar{\gamma}/\abs{Z}^2.
\end{equation}
Coherent and dissipative processes are encoded, respectively, in the real and
imaginary parts of the renormalized coefficients $K = A + i D$, $u = \lambda + i
\kappa$, and $u' = \lambda' + i \kappa'$.

We define the first three dimensionless scaling variables to be the ratios of
coherent to dissipative coefficients: $r_K = A/D$, $r_u = \lambda/\kappa$, and
$r_{u'} = \lambda'/\kappa'$.  Another three dimensionless variables are defined
by rescaling the loss coefficients $\kappa$ and $\kappa'$ and the condensate
density $\rho_0$:
\begin{equation}
  \label{eq:6}
  w = \frac{2 \kappa \rho_0}{\L^2 D}, \quad
  \tilde{\kappa} = \frac{\gamma \kappa}{2 \L D^2}, \quad
  \tilde{\kappa}' = \frac{\gamma^2 \kappa'}{4 D^3}.
\end{equation}
The flow equations for the couplings $\mathbf{r} = \left( r_K,r_u,r_{u'}
\right)^T$ and $\mathbf{s} = \left( w,\tilde{\kappa},\tilde{\kappa}' \right)^T$
form a closed set,
\begin{equation}
  \label{eq:7}
  \Lambda \partial_{\Lambda} \mathbf{r} = \beta_{\mathbf{r}}(\mathbf{r},\mathbf{s}), \quad
  \Lambda \partial_{\Lambda} \mathbf{s} = \beta_{\mathbf{s}}(\mathbf{r},\mathbf{s})
\end{equation}
(see SI for the explicit form). As a consequence of the
transformations~\eqref{eq:1} and~\eqref{eq:6}, these $\b$-functions acquire a
contribution from the running anomalous dimensions $\eta_a({\mathbf r},{\mathbf
  s}) = - \Lambda \partial_{\Lambda} \ln a$ associated with $a = Z, D, \gamma$.
\begin{figure}[b]
  \includegraphics{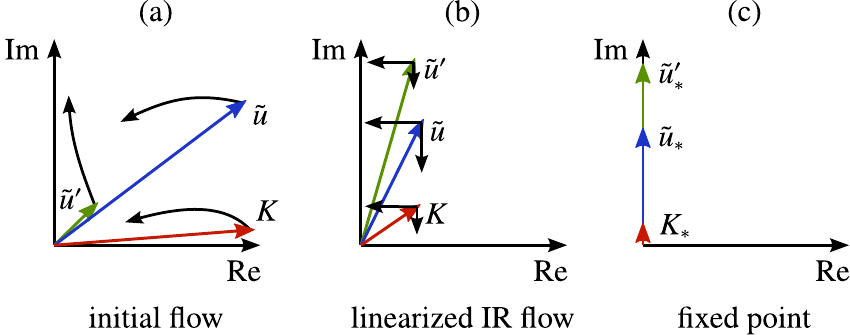}
  \caption{Flow in the complex plane of dimensionless renormalized
    couplings. (a) The microscopic action determines the initial values of the
    flow. Typically, the coherent propagation will dominate over the diffusion,
    $A \gg D$, while two-body collisions and two-body loss are on the same order
    of magnitude, $\tilde{\lambda} \approx \tilde{\kappa}$, with a similar
    relation for the marginal complex coupling $\tilde{u}'$. The initial flow is
    non-universal. (b) At criticality, the infrared (IR) flow approaches a
    universal linear domain encoding the critical exponents and anomalous
    dimensions. In particular, this regime is independent of the precise
    microscopic initial conditions. (c) The Wilson-Fisher fixed point describing
    the interacting critical system is purely imaginary.}
  \label{fig:flow}
\end{figure}

\emph{Critical properties --} The universal behavior near the critical point is
controlled by the infrared flow to a Wilson-Fisher like fixed point. The values
of the coupling constants at the fixed point, determined by solving $\b_{\bf
  s}(\br_{*},\bs_{*})=0$ and $\b_{\bf r}(\br_{*},\bs_{*})=0$, are given by: 
\begin{equation}
  \label{eq:3}
  \begin{split}
    \mathbf{r}_{*} & = \left( r_{K *},r_{u *},r_{u' *} \right) = \mathbf{0},\\
  \mathbf{s}_{*} & = \left( w_{*},\tilde{\kappa}_{*},\tilde{\kappa}'_{*}
  \right)\approx \left( 0.475,5.308,51.383 \right).
  \end{split}
\end{equation}
The fact that $\mathbf{r}_*=0$ implies that the fixed point action is purely
imaginary (or dissipative), as in Model A of Hohenberg and
Halperin~\cite{hohenberg77}, cf.~Fig.~\ref{fig:flow} (c). We interpret the fact
that the ratios of coherent vs.~dissipative couplings are zero at the fixed
point as a manifestation of decoherence at low frequencies in an RG
framework. The coupling values $\bs_{*}$ are identical to those obtained in an
equilibrium classical $O(2)$ model from functional RG calculations at the same
level of truncation~\cite{berges02}.

Let us turn to the linearized flow, which determines the universal behavior in
the vicinity of the fixed point. We find that the two sectors corresponding to
$\bs$ and $\br$ decouple in this regime, giving rise to a block diagonal
stability matrix
\begin{equation}
  {\partial\over \partial\ln \L} 
  \begin{pmatrix}
    \delta \mathbf{r} \\ \delta \mathbf{s}
  \end{pmatrix}
  =
  \begin{pmatrix}    
    N & 0 \\
    0 & S
  \end{pmatrix}
  \begin{pmatrix}
   \delta \mathbf{r} \\ \delta \mathbf{s}
  \end{pmatrix},
\end{equation}
where $\delta \mathbf{r} \equiv \mathbf{r}$, $\delta \mathbf{s} \equiv
\mathbf{s} - \mathbf{s}_{*}$, and $N,S$ are $3\times 3$ matrices (see SI).

The anomalous dimensions entering this flow are found by plugging the fixed
point values $\br_{*}, \bs_{*}$ into the expressions for $\eta_a(\br,\bs)$. We
obtain the scaling relation between the anomalous dimensions $\eta_Z=\eta_{\bar
  \g}$, valid in the universal infrared regime. This leads to cancellation of
$\eta_Z$ with $\eta_{\bar \g}$ in the static sector $S$ (see SI). The critical
properties in this sector, encoded in the eigenvalues of $S$, become identical
to those of the standard $O(2)$ transition. This includes the correlation length
exponent $\nu \approx 0.716$ and the anomalous dimension $\eta \approx 0.039$
associated with the bare kinetic coefficient $\bar{K}$. These values are in good
agreement with more sophisticated approximations~\cite{guida98}.

The equilibrium-like behavior in the $S$ sector can be seen as a result of an
emergent symmetry. Locking of the noise to the dynamical term implied by $\eta_Z
= \eta_{\bar \gamma}$ leads to invariance of the long wavelength effective
action (times $i$) under the transformation $\Phi_c(t,\mathbf{x}) \to \Phi_c(-
t,\mathbf{x}), \Phi_q(t,\mathbf{x}) \to \Phi_q(- t,\mathbf{x}) +
\frac{2}{\gamma} \sigma^z \partial_t \Phi_c(- t,\mathbf{x}), i \to - i$ with
$\Phi_\nu = (\phi_\nu, \phi_\nu^*)^T, \nu = (c,q)$, $\sigma^z$ the Pauli
matrix. It generalizes the symmetry noted in Refs.~\cite{aron10,canet11} to
models that include also reversible couplings. The presence of this symmetry
implies a classical FDR with a distribution function $F = 2T_\text{eff}/\omega$,
governed by an effective temperature $T_\text{eff} = \bar\gamma/ (4
\abs{Z})$. This quantity becomes scale independent in the universal critical
regime where $\bar \gamma \sim k^{- \eta_{\bar{\gamma}}}$ and $Z \sim
k^{-\eta_Z}$ cancel. We interpret this finding as an asymptotic low-frequency
thermalization mechanism of the driven system at criticality. The thermalized
regime sets in below the Ginzburg scale where fluctuations start to dominate,
for which we estimate perturbatively $\chi_G = \left( \gamma \kappa
\right)^2/\left( 16 \pi^2 D^3 \right)$ (see SI). The values entering here are
determined on the mesoscopic scale, and we specify them for exciton-polariton
systems in the SI based on Ref.~\cite{wouters10}. Above the scale $\chi_G$, no
global (scale independent) temperature can be defined in general. We note that,
unlike Hohenberg-Halperin type models, here the symmetry implied by $\eta_Z =
\eta_{\bar\gamma}$ is not imposed at the microscopic level of the theory, but
rather is emergent at the critical point.

The key new element in the driven-dissipative dynamics is encoded in the
decoupled ``drive'' sector (the $3 \times 3$ matrix $N$ in our case). It
describes the flow towards the emergent purely dissipative Model A fixed point
(see Fig.~\ref{fig:flow} (b)) and thus reflects a mechanism of low frequency
decoherence. This sector has no counterpart in the standard framework of
dynamical critical phenomena and is special to driven-dissipative systems. In
the deep infrared regime, only the lowest eigenvalue of this matrix governs the
flow of the ratios. This means that only one new critical exponent $\eta_r
\approx - 0.101$ is encoded in this sector. Just as the dynamical critical
exponent $z$ is independent of the static ones, the block diagonal structure of
the stability matrix ensures that the drive exponent is independent of the
exponents of the other sectors.

The fact that the inverse Green's function in Eq.~\eqref{eq:ansatz} is specified
by three real parameters, $\Re \bar{K}, \Im \bar{K}$, and $|Z|$ (the phase of
$Z$ can be absorbed by a $U(1)$ transformation) allows for only three
independent anomalous dimensions: $\eta_D$, $\eta_Z$ and the new exponent
$\eta_r$. Hence the extension of critical dynamics described here is
\emph{maximal}, i.e., no further independent exponent will be found. Moreover
this extension of the purely relaxational (Model A) dynamics leads to different
universality than an extension that adds reversible couplings compatible with
relaxation towards a Gibbs ensemble. The latter is obtained by adding real
couplings to the imaginary ones with the same ratio of real to imaginary parts
for all couplings~\cite{graham73,deker75,tauber01,longpaper}; in this case the
above symmetry is present, while absent in the general non-equilibrium case. The
compatible extension adds only an independent $1 \times 1$ sector $N$ to the
purely relaxational problem, for which we find $\eta_R = - 0.143 \neq \eta_r$.
This proves that the independence of dissipative and coherent dynamics defines
indeed a new non-equilibrium universality class with no equilibrium
counterpart. It is rooted in different symmetry properties of equilibrium
vs.~non-equilibrium situation.

\emph{Experimental detection} -- The novel anomalous dimension identified here
leaves a clear fingerprint in single-particle observables accessible with
current experimental technologies on different platforms. For ultracold atomic
systems this can be achieved via RF-spectroscopy~\cite{Stewart2008} close to the
driven-dissipative BEC transition. In exciton-polariton condensates, the
dispersion relation can be obtained from the energy- and
momentum resolved luminescence spectrum as
demonstrated in~\cite{Utsunomiya2008}. Using the RG scaling behavior of the
diffusion and propagation coefficients $D \sim D_0 \Lambda^{- \eta_D}$, $A = D
r_K \sim A_0 \Lambda^{- \eta_r - \eta_D}$, we obtain the anomalous scaling of
the frequency and momentum resolved, renormalized retarded Green's function $G^R
(\omega,\mathbf{q}) = (\omega - A_0 \abs{\mathbf{q}}^{2 - \eta_r -\eta_D} + i
D_0 \abs{\mathbf{q}}^{2 -\eta_D})^{-1}$, with $A_0$ and $D_0$ non-universal
constants. Peak position and width are implied by the complex dispersion $\omega
\approx A_0 \abs{\mathbf{q}}^{2.22} - i D_0 \abs{\mathbf{q}}^{2.12} $. The
energy resolution necessary to probe the critical behavior is again set by the
Ginzburg scale $\chi_G$ (see above).

\emph{Conclusions --} We have developed a Keldysh field theoretical approach to
characterize the critical behavior of driven-dissipative three dimensional Bose
systems at the condensation transition.  The main result presents a hierarchical
extension of classical critical phenomena. First, all static aspects are
identical to the classical $O(2)$ critical point. In the next shell of the
hierarchy a sub-class of the dynamical phenomena is identical to the purely
dissipative Model A dynamics of the equilibrium critical point. Finally we
identify manifestly non-equilibrium features of the critical dynamics, encoded
in a new independent critical exponent that betrays the driven nature of the
system.

\emph{Acknowledgements} -- We thank J. Berges, M. Buchhold, I. Carusotto,
T. Esslinger, T. Gasenzer, A. Imamoglu, J. M. Pawlowski, P. Strack, S. Takei,
U. C. T\"auber, C. Wetterich and P. Zoller for useful discussions.  This
research was supported by the Austrian Science Fund (FWF) through the START
grant Y 581-N16 and the SFB FoQuS (FWF Project No.
F4006-N16). 

\end{document}


\title{Supplementary Information for ``Dynamical Critical Phenomena in Driven-Dissipative Systems''}

\author{L. M. Sieberer$^{1,2}$, S. D. Huber$^{3,4}$, E. Altman$^{4,5}$, and S. Diehl$^{1,2}$
  \\{$^1$\small \em  Institute for Theoretical Physics, University of Innsbruck, A-6020 Innsbruck, Austria}\\
  {$^2$\small \em Institute for Quantum Optics and Quantum Information of the Austrian Academy of Sciences, A-6020 Innsbruck, Austria}\\
  {$^3$\small \em Theoretische Physik, Wolfgang-Pauli-Strasse 27, ETH Zurich, CH-8093 Zurich, Switzerland}\\
  {$^4$\small \em Department of Condensed Matter Physics, Weizmann Institute of Science, Rehovot 76100, Israel}\\
  {$^5$\small \em Department of Physics, University of California, Berkeley, CA 94720, USA}\\
}
\maketitle

\section{Open system dynamics}

Open system dynamics with local particle loss and gain can be modeled microscopically
by a many-body quantum master equation ($\hbar = 1$)
\begin{equation}
  \label{eq:meq}
  \partial_t \hat{\rho} = - i \left[ \hat{H},\hat{\rho} \right] + \mathcal{L} [\hat{\rho}].
\end{equation}
The dynamics of the system density matrix $\hat{\rho}$ has both a coherent
contribution due to the standard Hamiltonian for bosons of mass $m$
($\int_{\mathbf{x}} = \int d^3 \mathbf{x}$)
\begin{equation}  
  \hat{H} = \int_{\mathbf{x}} \hat{\psi}^{\dagger}(\mathbf{x}) \left( -
    \frac{\Delta}{2 m} - \mu \right) \hat{\psi}(\mathbf{x}) + \lambda
  \int_{\mathbf{x}} \hat{\psi}^{\dagger}(\mathbf{x})^2 \hat{\psi}(\mathbf{x})^2,
\end{equation}
and a dissipative one that is incorporated by the Liouville operator
\begin{multline}  
  \mathcal{L}[\hat{\rho}] = \gamma_p \int_{\mathbf{x}} \left(
    \hat{\psi}^{\dagger}(\mathbf{x}) \hat{\rho} \hat{\psi}(\mathbf{x}) -
    \frac{1}{2} \left\{ \hat{\psi}(\mathbf{x})
      \hat{\psi}^{\dagger}(\mathbf{x}), \hat{\rho} \right\} \right) \\
  + \gamma_l \int_{\mathbf{x}} \left( \hat{\psi}(\mathbf{x}) \hat{\rho}
    \hat{\psi}^{\dagger}(\mathbf{x}) - \frac{1}{2} \left\{
      \hat{\psi}^{\dagger}(\mathbf{x})
      \hat{\psi}(\mathbf{x}) ,\hat{\rho} \right\} \right) \\
  + 2 \kappa \int_{\mathbf{x}} \left( \hat{\psi}(\mathbf{x})^2 \hat{\rho}
    \hat{\psi}^{\dagger}(\mathbf{x})^2 - \frac{1}{2} \left\{
      \hat{\psi}^{\dagger}(\mathbf{x})^2 \hat{\psi}(\mathbf{x})^2, \hat{\rho}
    \right\} \right).
\end{multline}
Local Lindblad operators $\hat{\psi}^{\dagger}(\mathbf{x})$ and
$\hat{\psi}(\mathbf{x})$, respectively, correspond to the processes of
incoherent pumping and loss of single particles; $\hat{\psi}(\mathbf{x})^2$
describes the simultaneous loss of two particles. Associated rates are
$\gamma_p, \gamma_l$, and $2 \kappa$.

The investigation of critical phenomena at the stationary state phase transition
exhibited by this model is facilitated by a formulation in terms of a Keldysh
partition function~\cite{kamenev09:_keldy,altlandsimons} $\mathcal{Z} = \int
\mathcal{D} \psi_{+} \mathcal{D} \psi_{-} \, e^{i \mathcal{S}}$, which can be
subject to renormalization group methods. This partition function is fully
equivalent to the master equation~\eqref{eq:meq} and defined in terms of a
Keldysh action $\mathcal{S} = \mathcal{S}_H + \mathcal{S}_D$ with two
contributions corresponding to the commutator with the Hamiltonian (from now on
we will be using units such that $2 m = 1$; $\int_{t,\mathbf{x}} = \int d t \int
d \mathbf{x}$),
\begin{equation}
  \mathcal{S}_H = \sum_{\sigma = \pm} \sigma \int_{t,\mathbf{x}} \left[ \psi_{\sigma}^{*} \left(
      i \partial_t + \Delta + \mu \right) \psi_{\sigma} -
    \lambda \left( \psi_{\sigma}^{*} \psi_{\sigma} \right)^2 \right],
\end{equation}
and the dissipative Liouvillian,
\begin{multline}
  \mathcal{S}_D = - i \gamma_p \int_{t,\mathbf{x}} \left[ \psi_{+}^{*} \psi_{-}
    - \frac{1}{2} \left( \psi_{+} \psi_{+}^{*} + \psi_{-} \psi_{-}^{*} \right)
  \right] \\ - i \gamma_l \int_{t,\mathbf{x}} \left[ \psi_{+} \psi_{-}^{*} -
    \frac{1}{2} \left( \psi_{+}^{*} \psi_{+} + \psi_{-}^{*} \psi_{-} \right)
  \right] \\ - i 2 \kappa \int_{t,\mathbf{x}} \left\{ \left( \psi_{+}
      \psi_{-}^{*} \right)^2 - \frac{1}{2} \left[ \left( \psi_{+}^{*} \psi_{+}
      \right)^2 + \left( \psi_{-}^{*} \psi_{-} \right)^2 \right] \right\}.
\end{multline}
Expressing the Keldysh action in terms of classical and quantum fields, which
are defined as
\begin{equation}  
  \phi_c = \frac{1}{\sqrt{2}} \left( \psi_{+} + \psi_{-} \right), \quad
  \phi_q = \frac{1}{\sqrt{2}} \left( \psi_{+} - \psi_{-} \right),
\end{equation}
we recover Eq.~(2) of the main text.

\section{Functional Renormalization group equation}

Our approach to studying critical dynamics is based on the Wetterich functional
renormalization group~\cite{wetterich93} adapted to the Keldysh
framework (see~\cite{berges02,salmhofer01,pawlowski07,delamotte07,rosten12,boettcher12}
for reviews on the equilibrium formulation). Central to this method is the
functional $\Gamma_{\Lambda}[\phi_c,\phi_q]$ defined by~\cite{footnote_effact}
\begin{equation}
  \label{eq:effact}
  e^{i \Gamma_{\Lambda}[\phi_c,\phi_q]} = \int \mathcal D \delta\phi_c\mathcal
  D \delta\phi_q \, e^{i \mathcal{S}[\phi_c + \delta \phi_c,\phi_q
    +\delta\phi_q]+ i \Delta \mathcal{S}_{\Lambda}[\delta \phi_c,\delta \phi_q]}.
\end{equation}
Here $\Delta \mathcal{S}_{\Lambda} $ is a regulator function which suppresses
contributions to the above path integral from modes with spatial wave-vector
below the running cutoff $\Lambda$. Thus $\Gamma_{\Lambda}$ interpolates between
the classical action $\mathcal{S}$, when $\Lambda$ equals the UV cutoff $\Lambda_0$, and
the effective action functional $\Gamma[\phi_c,\phi_q]$~\cite{amitbook} when $\Lambda
\to 0$. The latter includes the effects of fluctuations on all scales. The
equation
\begin{equation}\label{eq:frg} 
  \partial_{\Lambda} \Gamma_{\Lambda} = \frac{i}{2} \Tr \left[ \left( \Gamma^{(2)}_{\Lambda}
      + R_{\Lambda} \right)^{-1} \partial_{\Lambda} R_{\Lambda} \right]
\end{equation}
describes the flow of the interpolating functional as a function of the running
cutoff $\Lambda$. In the following sections we first discuss the objects
appearing in~\eqref{eq:frg}, namely the second functional derivative
$\Gamma_{\Lambda}^{(2)}$ and the cutoff function $R_{\Lambda}$. Then we explain
how a closed set of flow equations for a finite number of coupling constants can
be obtained from the functional flow equation. Finally we detail the linearized
equations for the infrared flow to the Wilson-Fisher fixed point from which the
critical properties are inferred.

In suitable truncation schemes, results from high order epsilon expansion can be
reproduced from the exact flow equation~\eqref{eq:frg}. In our practical
calculation, we approach the critical point from the ordered phase. This allows
us to calculate the anomalous dimensions at one-loop order, due to the presence
of a finite condensate during the flow. Results obtained in this way have proven
to be competitive with high-order epsilon expansion or Monte Carlo simulations,
as referenced in the main text.

\section{The second variational derivative}

The second variation $\Gamma_{\Lambda}^{(2)}$ with respect to the fields is the
full inverse Green's function at the scale $\Lambda$, which in the case of an
interacting theory is field dependent. Practically we work in a basis of real
fields, related to the complex fields by
\begin{equation}
  \begin{pmatrix}
    \chi_{\nu,1}(Q) \\
    \chi_{\nu,2}(Q) 
  \end{pmatrix}
  = \frac{1}{\sqrt{2}}
  \begin{pmatrix}
    1  & 1 \\
    -i & i 
  \end{pmatrix}  
  \begin{pmatrix}
    \phi_{\nu}(Q) \\
    \phi^*_{\nu}(-Q) 
  \end{pmatrix},
\end{equation}
where $\nu = c, q$ is the Keldysh index. We gather the resulting four
independent field components in a field vector,
\begin{equation}
  \chi(Q) = \left( \chi_{c,1}(Q), \chi_{c,2}(Q), \chi_{q,1}(Q),
    \chi_{q,2}(Q) \right)^T.
\end{equation}
In this basis, $\Gamma_{\Lambda}^{(2)}$ is defined as
\begin{equation}
  \label{eq:6}
  \left( \Gamma_{\Lambda}^{(2)} \right)_{ij}(Q,Q') = \frac{\delta^2 \Gamma_{\Lambda}}{\delta
    \chi_i(-Q) \delta \chi_j(Q')},
\end{equation}
which is a matrix in the discrete field index $i = 1,2,3,4$ and in the
continuous momentum variable $Q = (\omega,\mathbf{q})$ collecting frequency and
spatial momentum. Accordingly, the trace in~\eqref{eq:frg} involves both an
integration over momenta and a sum over internal indices.

$\Gamma_{\Lambda}^{(2)}(Q,Q')$ is conveniently decomposed into a constant part
and a fluctuation part. The latter is a polynomial in momentum-dependent fields
and, therefore, a non-diagonal matrix in momentum space. In contrast, the
constant part is obtained by (i) inserting spatially constant field
configurations, i.e., $\chi(Q) = \chi \delta(Q)$ in momentum space, and (ii)
evaluating them at their stationary state values in the ordered phase. These
read
\begin{equation}\label{eq:backgr}
  \chi(Q)\bigr\rvert_{\mathrm{ss}} = \left( \sqrt{2 \rho_0}, 0,0,0 \right)^T \delta(Q).
\end{equation}
(Without loss of generality we choose the condensate amplitude to be real.) As
a result, the constant part is diagonal in momentum space,
\begin{equation}  
  P_{\Lambda}(Q) \delta(Q - Q') \equiv \Gamma_{\Lambda}^{(2)}(Q,Q')
  \bigr\rvert_{\mathrm{ss}},
\end{equation}
and is structured into retarded, advanced, and Keldysh blocks,
\begin{equation}
  \label{eq:4}
  P_{\Lambda}(Q) =
  \begin{pmatrix}
    0 & P^A(Q) \\
    P^R(Q) & P^K
  \end{pmatrix}.
\end{equation}
(For notational simplicity, we suppress the scale index $\Lambda$ for the
different blocks and their respective entries.) The retarded and advanced blocks
are mutually hermitian conjugate (we decompose $Z$ and $\bar{K}$ into real and
imaginary parts, $Z = Z_R + i Z_I$, $\bar{K} = \bar{A} + i \bar{D}$),
\begin{eqnarray}
  \label{eq:13}
  P^R(Q) &=&
  \begin{pmatrix}
    - i Z_I \omega - \bar{A} \mathbf{q}^2 - 2 \Re(\bar{u}) \rho_0 & i Z_R
    \omega - \bar{D} \mathbf{q}^2 \\
    - i Z_R \omega + \bar{D} \mathbf{q}^2 + 2 \Im(\bar{u}) \rho_0 & - i Z_I
    \omega - \bar{A} \mathbf{q}^2
  \end{pmatrix}, \nonumber \\
  P^A(Q) &=& \left( P^R(Q) \right)^{\dagger}.
\end{eqnarray}
Note that $\det P^R (Q=0) = \det P^A (Q=0) =0$; the existence of a gapless mode
associated to the broken $U(1)$ symmetry is thus ensured in our truncation at
all scales $\Lambda$. For the Keldysh block we have
\begin{equation}
  P^K = i \bar{\gamma} \id.
\end{equation}

\section{The regulator function}

The cutoff contribution $\Delta \mathcal{S}_{\Lambda}$ is used in Eq.~\eqref{eq:effact} to
generate the effective action $\Gamma_{\Lambda}$ from the microscopic action
$\mathcal{S}$ by suppressing contributions from momenta below $\Lambda$. Its
second functional derivative $R_{\Lambda}=\Delta \mathcal{S}_{\Lambda}^{(2)}$
enters the exact flow equation~\eqref{eq:frg}. We choose an optimized cutoff
function~\cite{litim00} of the form
\begin{equation}
  \label{eq:7}
  R_{\Lambda}(Q) = \left( \mathbf{q}^2 - \Lambda^2 \right) \theta(\Lambda^2
  -
  \mathbf{q}^2)
  \begin{pmatrix}
    0 & R^R \\
    R^A & 0
  \end{pmatrix},
\end{equation}
where
\begin{equation}
  R^R =
  \begin{pmatrix}
    - \bar{A} & - \bar{D} \\
    \bar{D} & - \bar{A}
  \end{pmatrix}, \quad R^A = \left( R^R \right)^T.
\end{equation}
Due to the $\theta$-function in~\eqref{eq:7}, in the regularized inverse Green's
function
\begin{equation}
  G_{\Lambda}^{-1} = P_{\Lambda} + R_{\Lambda},
\end{equation}
momenta $\mathbf{q}^2$ smaller than the running scale $\Lambda^2$ acquire an
effective mass $\propto \Lambda^2$ and we have $\det G_{\Lambda}^{-1}(Q = 0)
\neq 0$, which ensures that momentum integrals over Green's functions are
infrared convergent. Note that it is sufficient for $R_{\Lambda}$ to modify only
the retarded and advanced blocks (i.e., the spectrum) of the inverse
Green's function. The choice of a frequency-independent cutoff allows us to perform
frequency integrals analytically.

The interpolation property of $\Gamma_{\Lambda}$ between the classical action
$\mathcal{S}$ and the effective action $\Gamma$ is guaranteed by the limiting
behavior \cite{berges09}
\begin{equation}
  \lim_{\Lambda^2 \to \Lambda_0^2} R_{\Lambda} \sim
  \Lambda_0^2, \quad \lim_{\Lambda^2 \to 0} R_{\Lambda} = 0.
\end{equation}

\section{Flow of the effective potential}

In equilibrium problems, an important object for practical calculations is the
effective potential. It describes the homogeneous part of the effective action
and is obtained by evaluating the full effective action at spatially homogeneous
field configurations, $\bar U = \Gamma/\Omega \bigr\rvert_{\chi(Q) =
  \chi\delta(Q)}$ ($\Omega$ is the quantization volume). In the framework of a
derivative expansion, a closed flow equation can be derived for this object,
which serves as a compact generating functional for the flow of all local
couplings to arbitrarily high order. Here we provide the Keldysh analog of this
construction, where the key difference roots in the occurrence of two field
variables $\phi_c,\phi_q$, in contrast to a single field in
equilibrium. However, for a theory which obeys the power counting discussed in
the main text, we can parameterize the homogeneous part of the effective action
as
\begin{equation}
  \bar{V} = \frac{\partial \bar{U}}{\partial
    \phi_c} \phi_q + \frac{\partial \bar{U}^{*}}{\partial \phi_c^{*}} \phi_q^{*} + i
  \bar{\gamma}  \phi_q^{*}  \phi_q,
\end{equation}
with $\bar U = \bar U (\phi^*_c\phi_c)$ dependent on the $U(1)$ invariant
combination of classical fields only, this function thus being the direct
counterpart of the effective potential.  A flow equation can be derived for the
auxiliary object $\bar V$, which reads (we introduce a dimensionless scale
derivative $ \dL \equiv \Lambda\partial_{\Lambda}$)
\begin{equation}
  \label{eq:2}
  \dL \bar{V}  = - \frac{i}{2} \int_Q \tr \left[
    \mathcal G_{\Lambda}(Q) \dL R_{\Lambda}(Q) \right].
\end{equation}
Here, the inverse of $\mathcal G_{\Lambda}$ is obtained from the full second
functional variation by evaluating it at homogeneous field configurations (step
(i) above Eq.~\eqref{eq:backgr}), however without inserting the stationary state
values (step (ii)): $\mathcal G^{-1}_{\Lambda} = \Gamma_{\Lambda}^{(2)}
\bigr\rvert_{\chi(Q) = \chi\delta(Q)} + R_{\Lambda}$. $\mathcal G_{\Lambda}$ is
then diagonal in momentum space, and so the trace in Eq.~\eqref{eq:frg} reduces
to a single momentum integration, giving rise to the above compact form. In
contrast to $G^{-1}_{\Lambda}$, $\mathcal G_{\Lambda}^{-1}$ has a non-vanishing
upper left block $P^H$. However, it vanishes when the background fields are set
to their stationary state values, $P^H \bigr\rvert_{\mathrm{ss}} = 0,$ which is
a manifestation of causality in the Keldysh formalism
\cite{kamenev09:_keldy,altlandsimons}.

From this equation we obtain the $\beta$-functions for the momentum-independent
couplings by evaluating appropriate derivatives with respect to the $U(1)$
invariants
\begin{equation}
  \rho_c = \phi_c^{*} \phi_c, \quad \rho_{cq} = \phi_c^{*} \phi_q =
  \rho_{qc}^{*}, \quad \rho_q = \phi_q^{*} \phi_q.
\end{equation}
at their stationary state values $\rho_c\rvert_{\mathrm{ss}} = \rho_0$,
$\rho_{cq}\rvert_{\mathrm{ss}} = \rho_{qc}\rvert_{\mathrm{ss}} =
\rho_q\rvert_{\mathrm{ss}} = 0$. Specifically, we use the projection
prescriptions
\begin{equation}
  \label{eq:5}
  \begin{split}
    \dL \rho_0 & = \beta_{\rho_0} = -
    \frac{1}{u} \left[ \partial_{\rho_{cq}}  \dL \bar{V} \right]_{\mathrm{ss}}, \\
    \dL \bar{u} & = \beta_{\bar{u}} = \bar{u}'
    \dL \rho_0 + \left[ \partial_{\rho_c \rho_{cq}}^2  \dL \bar{V} \right]_{\mathrm{ss}}, \\
    \dL \bar{u}' & = \beta_{\bar{u}'}
    = \left[ \partial_{\rho_c}^2 \partial_{\rho_{cq}}  \dL \bar{V} \right]_{\mathrm{ss}}, \\
    \dL \bar{\gamma} & = \beta_{\bar{\gamma}} = i \rho_0
    \left[ \partial_{\rho_{cq} \rho_{qc}}^2 \dL \bar{V} \right]_{\mathrm{ss}}.
  \end{split}
\end{equation}
Calculation of the explicit expressions here and below is largely automatized
using \mathematica.

\section{Flow of the inverse propagator}

While the flow equation for the effective potential~\eqref{eq:2} generates
$\beta$-functions for all momentum-independent couplings, the flow of the
complex dynamic $Z$ and kinetic $\bar{K}$ couplings, which constitute the
momentum-dependent part of the effective action~(3), is determined by the flow
equation for the inverse propagator. We obtain the latter by taking the second
variational derivative of the exact flow equation~\eqref{eq:frg} and setting the
background fields to their stationary state values Eq.~\eqref{eq:backgr},
\begin{multline}
  \label{eq:1}
  \dL P_{\Lambda,ij}(Q) = \\ \frac{i}{2}  \int_{Q'} \tr \left[
    G_{\Lambda}^2(Q'
    - Q) \dL R_{\Lambda}(Q' - Q) \gamma_i G_{\Lambda}(Q') \gamma_j \right. \\
  \left. + G_{\Lambda}(Q' - Q) \gamma_i G_{\Lambda}^2(Q') \dL R_{\Lambda}(Q')
    \gamma_j \right],
\end{multline}
where
\begin{equation}
  \gamma_{i,jl} \delta(P - P' + Q) = \frac{\delta
    \Gamma^{(2)}_{\Lambda,jl}(P,P')}{\delta \chi_i(Q)}
  \biggr\rvert_{\mathrm{ss}}.
\end{equation}
In Eq.~\eqref{eq:1} we omit tadpole contributions $\propto
\Gamma_{\Lambda}^{(4)}$, which do not depend on the external momentum $Q$ and
hence do not contribute to the flow of $Z$ or $\bar{K}$. For these we use the
projection prescriptions
\begin{equation}
  \label{eq:3}
  \begin{split}    
    \dL Z & = \beta_Z = - \frac{1}{2} \partial_{\omega} \tr \left[ \left( \id +
        \sigma_y \right) \dL P^R(Q) \right] \Bigr\rvert_{Q = 0}, \\
    \dL \bar{K} & = \beta_{\bar{K}} = \partial_{\mathbf{q}^2} \left[ \dL
      P^R_{22}(Q) + i \dL P^R_{12}(Q) \right]
    \Bigr\rvert_{Q = 0}.
  \end{split}
\end{equation}
The $\beta$-functions~\eqref{eq:5} and~\eqref{eq:3} constitute the components of
$\beta_{\mathbf{g}} = \left( \beta_Z, \beta_{\bar{K}}, \beta_{\rho_0},
  \beta_{\bar{u}}, \beta_{\bar{u}'}, \beta_{\bar{\gamma}} \right)^T$.

\section{Rescaled flow equations}

We write the flow equation for the complex dynamic coupling $Z$ in the form
\begin{equation}
  \label{eq:10}
  \dL Z = - \eta_Z Z.
\end{equation}
The anomalous dimension $\eta_Z$ is an algebraic function of the rescaled
couplings~(6) and $\rho_0$. The same applies to the $\beta$-functions of the
latter,
\begin{equation}
  \label{eq:8}
  \begin{split}
    \dL K & = \beta_K = \eta_Z K + \frac{1}{Z} \beta_{\bar{K}}, \\
    \dL u & = \beta_u = \eta_Z u + \frac{1}{Z} \beta_{\bar{u}}, \\
    \dL u' & = \beta_{u'} = \eta_Z u' + \frac{1}{Z} \beta_{\bar{u}'}, \\
    \dL \gamma & = \beta_{\gamma} = \left( \eta_Z + \eta_Z^{*} \right) \gamma +
    \frac{1}{\abs{Z}^2} \beta_{\bar{\gamma}}.
  \end{split}
\end{equation}
In particular, the very right expressions in these equations
($\beta_{\bar{K}}/Z$ etc.) are functions of the rescaled couplings alone. In
terms of these variables, therefore, all explicit reference to the running
coupling $Z$ is gone, and we have effectively traded the differential flow
equation for $Z$ for the algebraic expression for its anomalous dimension
$\eta_Z$.

All couplings except for $\gamma$ are complex valued. Taking real and imaginary
parts of the $\beta$-functions for $K$, $u$, and $u'$ yields the flow equations
for $A$, $D$, $\lambda$, $\kappa$, $\lambda'$, and $\kappa'$ respectively,
\begin{equation}
  \label{eq:15}
  \begin{aligned}
    \dL A & = \beta_A = \Re \beta_K, & \dL D & = \beta_D = \Im \beta_K, \\
    \dL \lambda & = \beta_{\lambda} = \Re \beta_u, & \dL \kappa & =
    \beta_{\kappa} = \Im \beta_u, \\
    \dL \lambda' & = \beta_{\lambda'} = \Re \beta_{u'}, & \dL \kappa' & =
    \beta_{\kappa'} = \Im \beta_{u'}. \\
  \end{aligned}
\end{equation}
The $\beta$-functions for the ratios $\mathbf{r} = \left( r_K, r_u, r_{u'}
\right)^T$ are then
\begin{equation}
  \label{eq:12}
  \begin{split}
    \dL r_K & = \beta_{r_K} = \frac{1}{D} \beta_A - \frac{r_K}{D} \beta_D, \\
    \dL r_u & = \beta_{r_u} = \frac{1}{\kappa} \beta_{\lambda} -
    \frac{r_u}{\kappa} \beta_{\kappa}, \\
    \dL r_{u'} & = \beta_{r_{u'}} = \frac{1}{\kappa'} \beta_{\lambda'} -
    \frac{r_{u'}}{\kappa'} \beta_{\kappa'}.
  \end{split}
\end{equation}
The number of flow equations can be further reduced by introducing anomalous
dimensions for $D$ and $\gamma$,
\begin{equation}
  \label{eq:9}
  \begin{split}
    \dL D & = - \eta_D D, \\
    \dL \gamma & = - \eta_{\gamma} \gamma.
  \end{split}
\end{equation}
As for the dynamic coupling $Z$ in terms of the rescaled variables $K$, $u$,
$u'$, $\gamma$ and $\rho_0$, all explicit reference to $D$ and $\gamma$ drops
out, and we obtain for the couplings $\mathbf{s} = \left( w, \tilde{\kappa},
  \tilde{\kappa}' \right)^T$ defined in Eq.~(5)
\begin{equation}
  \label{eq:11}
  \begin{split}
    \dL w & = \beta_w = - \left( 2 - \eta_D \right) w + \frac{w}{\kappa}
    \beta_{\kappa} + \frac{2 \kappa}{\Lambda^2 D} \beta_{\rho_0}, \\
    \dL \tilde{\kappa} & = \beta_{\tilde{\kappa}} = - \left( 1 - 2 \eta_D +
      \eta_{\gamma} \right) \tilde{\kappa} + \frac{\gamma}{2 \Lambda D^2}
    \beta_{\kappa}, \\ \dL \tilde{\kappa}' & = \beta_{\tilde{\kappa}'} = -
    \left( - 3 \eta_D + 2 \eta_{\gamma} \right) \tilde{\kappa}' +
    \frac{\gamma^2}{4 D^3} \beta_{\kappa'}.
  \end{split}
\end{equation}
In summary, the transformations~(4) and~(5) result in the closed system~(6) for
$\mathbf{r}$ and $\mathbf{s}$ with $\beta_{\mathbf{r}} = \left( \beta_{r_K},
  \beta_{r_u}, \beta_{r_{u'}} \right)^T$ given by Eq.~\eqref{eq:12} and
$\beta_{\mathbf{s}} = \left( \beta_w, \beta_{\tilde{\kappa}},
  \beta_{\tilde{\kappa}'} \right)^T$ given by Eq.~\eqref{eq:11}. The flows of
$Z$, $D$, and $\gamma$ are decoupled and determined by the anomalous
dimensions~\eqref{eq:10} and~\eqref{eq:9}, which are themselves functions of
$\mathbf{r}$ and $\mathbf{s}$.

\section{Critical properties}

For the analysis of critical behavior, we need to find a scaling solution to the
flow equations for the bare couplings or, equivalently, a fixed point
$\mathbf{r}_{*}$, $\mathbf{s}_{*}$ of the flow of dimensionless rescaled
couplings,
\begin{equation}
  \beta_{\mathbf{r}}(\mathbf{r}_{*},\mathbf{s}_{*}) =
  \beta_{\mathbf{s}}(\mathbf{r}_{*},\mathbf{s}_{*}) = \mathbf{0}.
\end{equation}
This non-linear algebraic set of equations has a non-trivial solution given by
Eq.~(7). In order to characterize the infrared flow in the vicinity of the fixed
point (encoding the critical exponents we are interested in here), we study the
flow of the couplings linearized around the fixed point, cf.\,Eq.~(8). The
stability matrices $N$ and $S$ in this equation read explicitly
\begin{gather}
  N = \nabla_{\mathbf{r}}^T \beta_{\mathbf{r}} \bigr\rvert_{\mathbf{r} =
    \mathbf{r}_{*}, \mathbf{s} = \mathbf{s}_{*}} =
  \begin{pmatrix}
    0.0525 & 0.0586 & 0.0317 \\
    -0.0002 & -0.0526 & 0.1956 \\    
    0.4976 & -2.3273 & 1.9725
  \end{pmatrix}, \\
  S = \nabla_{\mathbf{s}}^T \beta_{\mathbf{s}} \bigr\rvert_{\mathbf{r} =
    \mathbf{r}_{*}, \mathbf{s} = \mathbf{s}_{*}} =
  \begin{pmatrix}
    -1.6204 & 0.0881 & 0.0046 \\
    -3.1828 & 0.2899 & 0.0363 \\
    -15.3743 & -42.2487 & 2.1828
  \end{pmatrix},
\end{gather}
without coupling between $\mathbf r$ and $\mathbf s$ sectors. At present we
cannot rule out that an extended truncation would couple them. However, since we
already include all relevant and marginal couplings, we expect the decoupling to
be robust or at least approximately valid to a good accuracy.

The infrared flow of $Z$, $D$, and $\gamma$ is determined by the values of the
respective anomalous dimensions at the fixed point. Equations~\eqref{eq:10}
and~\eqref{eq:9} imply the scaling behavior
\begin{equation}
  Z \sim \Lambda^{- \eta_Z}, \quad D \sim \Lambda^{- \eta_D}, \quad \gamma \sim \Lambda^{- \eta_{\gamma}}
\end{equation}
for $\Lambda \to 0$. While $\eta_D$ and $\eta_{\gamma}$ describe the flow of
real quantities and are, therefore, themselves real by definition, $\eta_Z$ is
in general a complex valued function of $\mathbf{r}$ and $\mathbf{s}$. At the
fixed point, however, the imaginary part vanishes,
\begin{equation}
  \Im \eta_Z = 0,
\end{equation}
which ensures scale invariance of the full effective action at the critical
point.

As is indicated in the main text, the emergence of $O(2)$ model critical
properties in the sector $\mathbf{s}$ is due to the scaling relation $\eta_Z =
\eta_{\bar{\gamma}}$, which ensures that these anomalous dimensions compensate
each other in the $\beta$-functions for the couplings $\mathbf{s}$.   (The anomalous dimensions $\eta_{\bar{\gamma}}$ and $\eta_{\gamma}$ associated
  with the bare and renormalized noise vertices, respectively, are related via
  $\eta_{\gamma} = \eta_{\bar{\gamma}} - 2 \Re \eta_Z$, as follows from Eq.~(4)
  in the main text.) This can be seen most simply by expressing, e.g.,
$\tilde{\kappa}$ in terms of bare quantities,
\begin{equation}
  \label{eq:16}
  \tilde{\kappa} = \frac{\gamma \Im(u)}{2 \Lambda \Im(K)^2} = \frac{\gamma
    \Im(\bar{u}/Z)}{2 \Lambda \Im(\bar{K}/Z)^2}.
\end{equation}
In this form it is apparent that the scaling $\sim \Lambda^{- \eta_Z}$ which
applies to both $Z$ and $1/\gamma$ drops out. Similar arguments hold for $w$ and
$\tilde{\kappa}'$. Alternatively, the cancellation of $\eta_Z$ and
$\eta_{\gamma}$ in the $\beta$-functions can be seen explicitly by inserting
Eqs.~\eqref{eq:8} and~\eqref{eq:15} in \eqref{eq:11}. What remains is a
dependence on $\eta \equiv \eta_D + \eta_Z$ which is just the anomalous
dimension associated with the bare kinetic coefficient $\bar{K}$.

\section{Ginzburg Criterion} 

We estimate the extent of the universal critical domain governed by the
linearized regime of the Wilson-Fisher fixed point, which provides us with an
estimate of both the extent of the thermalized regime as well as the energy
resolution necessary to probe the critical behavior. This is done by calculating
the Ginzburg scale, i.e., the distance from the phase transition where
fluctuations on top of the quadratic Bogoliubov-type theory become dominant~\cite{amitbook}: We equate the bare distance from
the phase transition $\chi$ to the corresponding one-loop correction, yielding
\begin{equation}
  \chi_G = \frac{1}{D^3} \left( \frac{\gamma \kappa}{4 \pi} \right)^2.
\end{equation}
Here, the parameters $\gamma, \kappa$, and $D$ are those appearing in the
mesoscopic description of the system. In the case of exciton-polariton
condensates, $\gamma$ and $\kappa$ can thus be read off from the dGPE and the
noise correlator~\cite{carusotto2005}. The parameter $D$ typically does not
appear explicitly in this description. However, it is included effectively in a
complex prefactor of the time derivative in the dGPE ($m_{\mathit{LP}}$ is the
mass of the lower polariton)
\begin{equation}
  \label{eq:18}
  i \left( 1 + i \Delta Z
  \right) \partial_t \psi = \left( - \frac{1}{2 m_{\mathit{LP}}} \nabla^2 + \dotsb
  \right) \psi.
\end{equation}
Such a term results from two physical mechanisms. First, it describes the leading
frequency dependence of the pumping process~\cite{wouters10_PRL}. To account for
this effect, a convenient parameterization is $\Delta Z = P/ \left( 2 \Omega_K
\right)$ which is proportional to the pumping strength $P$, and where $\Omega_K$
is the gain cutoff frequency. Second, it results from energy relaxation due to
scattering of the lower polaritons with high frequency photons and
excitons~\cite{wouters10_PRB}. These processes are captured by the form $\Delta
Z = \kappa \bar{n}/2$ scaling linearly with the time averaged density $\bar{n}$,
and a phenomenological relaxation constant $\kappa$.

Dividing the dGPE~\eqref{eq:18} by $1 + i \Delta Z$ leads to an effective
kinetic term $ - \tfrac{1 - i \Delta Z}{1 + \Delta Z^2} \tfrac{\nabla^2}{2
  m_{\mathit{LP}}} \psi$, resulting in a mesoscopic coherent propagation
coefficient $A = \frac{1}{1 + \Delta Z^2} \frac{1}{2 m_{\mathit{LP}}}$ and an
effective mesoscopic diffusion constant $D = \frac{\Delta Z}{1 + \Delta Z^2}
\frac{1}{2 m_{\mathit{LP}}}$ entering Eq.~(1) in the main text, and the above
Ginzburg criterion.

Finally, we would like to contrast the Ginzburg scale to a scale identified
in~\cite{carusotto2005,wouters10_PRL}. This scale indicates a crossover between
a sonic and a purely diffusive excitation spectrum within the symmetry broken
phase, and takes the value $\omega_c = \kappa\rho_0$.

The Ginzburg scale identifies the frequency scale below which critical
fluctuations become more dominant than the ``bare'' terms, which occur in
Bogoliubov theory. It is therefore only meaningful -- and makes a statement
about -- the physics close to the phase transition, where the order parameter
goes to zero.

Instead, the crossover scale in~\cite{carusotto2005,wouters10_PRL} is determined
within the symmetry broken phase and is obtained within Bogoliubov theory,
without the need of a calculation of fluctuation corrections. This is justified,
because within the symmetry broken phase there are no critical fluctuations and
mean field plus Bogoliubov theory are valid.

As implied by this comparison, these two scales are not directly related to each
other and address different physical questions. In particular, the crossover
scale in~\cite{carusotto2005,wouters10_PRL} tends to zero when approaching the
phase transition by construction, $\rho_0 \to 0$. Therefore, in the vicinity of
the critical point, the discussion of diffusive vs.\ coherent dynamics is a more
subtle issue. How it works quantitatively is addressed by the calculation of the
critical exponents, with the key finding of universal decoherence: The coherent
dynamics fades out faster than the dissipative one, measured by the critical
exponent $\eta_r$.